\documentclass[a4paper,11pt]{article}
\pdfoutput=1 

\usepackage{jheppub} 

\usepackage{def_simbol}

\title{\boldmath Intra-cone transition effect to magnetoconductivity in Dirac semimetal}

\author{Aya Kagimura, }
\author{Tetsuya Onogi}

\affiliation{Department of Physics, Osaka University, Toyonaka, Osaka 560-0043, Japan}

\emailAdd{kagimura@het.phys.sci.osaka-u.ac.jp}
\emailAdd{onogi@phys.sci.osaka-u.ac.jp}

\abstract{	We study the transport of the fermions with a small mass in the presence of Coulomb impurities, which could be realized in slightly distorted Dirac semimetals. Using the semiclassical Boltzmann equation, we derive the relaxation times for two kinds of intra-cone transition process. One is due to the effect of mass, and the other is due to the excited states in Landau levels under the weaker magnetic field. Hence we derive the mass dependence and the magnetic field dependence of the 
	longitudinal magnetoconductivity in the presence of parallel electric and magnetic fields. 
}

\begin{document}
\hspace{12cm} OU-HET-951
\maketitle
\flushbottom

\section{Introduction}\label{sec:Introduction}
Chiral anomaly
or
Adler-Bell-Jackiw anomaly\cite{PhysRev.177.2426,Bell1969},
which was discovered
in 1969, is an important concept in
gauge theories.
The physical consequence of the chiral
anomaly is that
massless fermions
coupled with
the
electromagnetic gauge field, the chiral fermion number
$N_5$ is not conserved
but
obeys the anomaly equation 
\begin{equation}
	\frac{dN_5}{dt}=\frac{e^2}{2\pi ^2}{\bf E\cdot B}	\label{1-anomaly equation}.
\end{equation}

In 1983 Nielsen and Ninomiya pointed out that condensed matter systems have an effect
which arises essentially from the same
mechanism of the chiral anomaly\cite{Nielsen:1983rb}. They considered a band structure with two Weyl nodes which possess the opposite chirality, applying the parallel electric and magnetic field. The fermion states are quantized by the magnetic field to form Landau levels, and the fermion gets drifted by the electric field within a given level. Then the equation for the chiral fermion number completely matches with Eq.(\ref{1-anomaly equation}). At the same time, drifted fermions get scattered back by impurities, acoustic phonons, or other electrons. The balance between the drift and the scattering determines the magnitude of the electric current measured in observation. They predicted the enhancement of the magnetoconductivity proportional to ${\bf E\cdot B}$ caused by the chiral anomaly effect. 

Their prediction had not been tested in the observation for 30 years, because the example of Weyl nodes had not been discovered in solids. The recent studies of the topological band structure and Berry curvature changed the situation. The interest of Weyl semimetals has begun with the theoretical proposal by Wan et al.\cite{PhysRevB.83.205101},
which has led to the study of various materials both in theories and experiments. 
From the theoretical studies, it was also found that the gapless level crossing\cite{PhysRevLett.108.140405} is protected by the crystal symmetry. This has led to the theoretical predictions\cite{PhysRevB.85.195320, PhysRevB.88.125427} and experimental observations of gapless three dimensional Dirac semimetals such as $Na_3Bi$ and $Cd_3Se_2$\cite{Liu864,2014NatCo...5E3786N,nmat3990,PhysRevLett.113.027603,2013arXiv1312.7624X}. There are several papers which report 
the unusual magnetoresistance in Weyl semimetals\cite{PhysRevX.5.031023, Zhang:2015gwa} and Dirac semimetals\cite{Li:2014bha, 2015arXiv150308179X, 2015arXiv150407698Z}. It would be interesting to compare the experimental observations to theoretical predictions.

To make a theoretical prediction at the quantitative level, one needs to know the relaxation time. Although there exists an early calculation in 1956\cite{PhysRev.104.900}, it is unsatisfactory since it was estimated for non-relativistic fermions in the quantum limit. Since the experiment measures the magnetoconductivity of semimetals for a wide range of the magnetic field strength ($0\sim 10$ Tesla), it is needed to predict the relaxaion time for electrons in the Dirac or Weyl semimetals with the magnetic field both in and away from the quantum limit. 

A theoretical prediction at the quantitative level is a rather difficult problem, since the relaxation occurs through transitions between Weyl (or Dirac) cones at different momentum points in the Brillouin zone (called as `inter-cone transition'), which is highly dependent on the material as well as its modeling. It would be nice if we can predict some universal features of the magnetoconductivity which is model independent.

The purpose of our study is to offer such a model independent theoretical prediction using the low energy effective theory. For this purpose, we focus on the effects on the magnetoconductivity for the Dirac semimetals due to the change of external parameters such as the magnetic field or the mass gap from the mechanical strain\cite{PhysRevB.84.085106}. We can expect that the change from the ideal magnetoresistance is triggered by the onset of `intra-cone' transition so that the effects can be described by the low energy effective theory for the single Dirac cone.  Using the action for the relativistic Dirac fermion as the low energy effective theory, we derive a general formula for the relaxation time due to impurities 
for the Dirac semimetal with a mass gap $m$ from the mechanical strain and  under the magnetic field $B$ including the regime away from the quantum limit. Using our formula,
we predict a drastic change in the magnetoconductivity. Although there has yet been no clear experimental observation of such effects, our prediction may offer a deeper understanding of the magnetoconductivity as well as
the interesting technological applications. 

In the theoretical development, the chiral anomaly contribution to the conductivity is discussed in a semiclassical argument\cite{PhysRevB.88.104412}. The phenomena that the imbalance of the chemical potential between the left- and right-handed mode induces the electric current under the magnetic field is now called the chiral magnetic effect and is now getting a renewed interest in the quark gluon plasma\cite{Fukushima:2008xe}, the electro-weak plasma in the early universe\cite{Joyce:1997uy}, and neutrinos in supernovae\cite{Yamamoto:2015gzz}. Our approach may also be extended to these systems. 

This paper is organized as follows. In Sec.\ref{sec:Basics}, we review the basics of the transport theory, and derive the relation between the electric current and the relaxation time. In Sec.\ref{sec:Relaxation}, we give the mass dependence of the relaxation time for the massive Dirac fermion. The magnetic field dependence of the magnetoconductivity in the weak magnetic field are shown in Sec.\ref{sec:Magnetoconductivity}. Finally, we summarize our study and give a discussion in Sec.\ref{sec:Summary}. 
\section{Basics of the transport theory}	\label{sec:Basics}
In this section, we review the basics of the transport theory and the relation between the electric current and the relaxation times when both the electric field ${\bf E}$ and the magnetic field ${\bf B}$ are applied along the $z$-axis. Following the discussion in Ref.\cite{Nielsen:1983rb}, we consider the relativistic fermion with a small mass.  Let us take the direction of the magnetic field as the $z$-axis. Due to the magnetic field, the electron states in the $x$ and $y$ directions form Landau levels. In the case that one is interested in the low energy physics under the strong magnetic field, the effective system becomes a $1+1$ dimensional electron system since only the lowest Landau level contribute to the physics. When the electric field in $z$ direction is applied, the electron gets the drifting force and the electric current flows. Due to the scattering from impurities, phonon excitations and other electrons, the momentum of the electrons are flipped and the current becomes static as a result of the balance between the drift and the relaxation by the scatterings. 

\begin{figure}[htbp]
	\begin{minipage}{0.5\hsize}
		\centering
		\includegraphics[bb= 0 0 496 328, width=8cm]{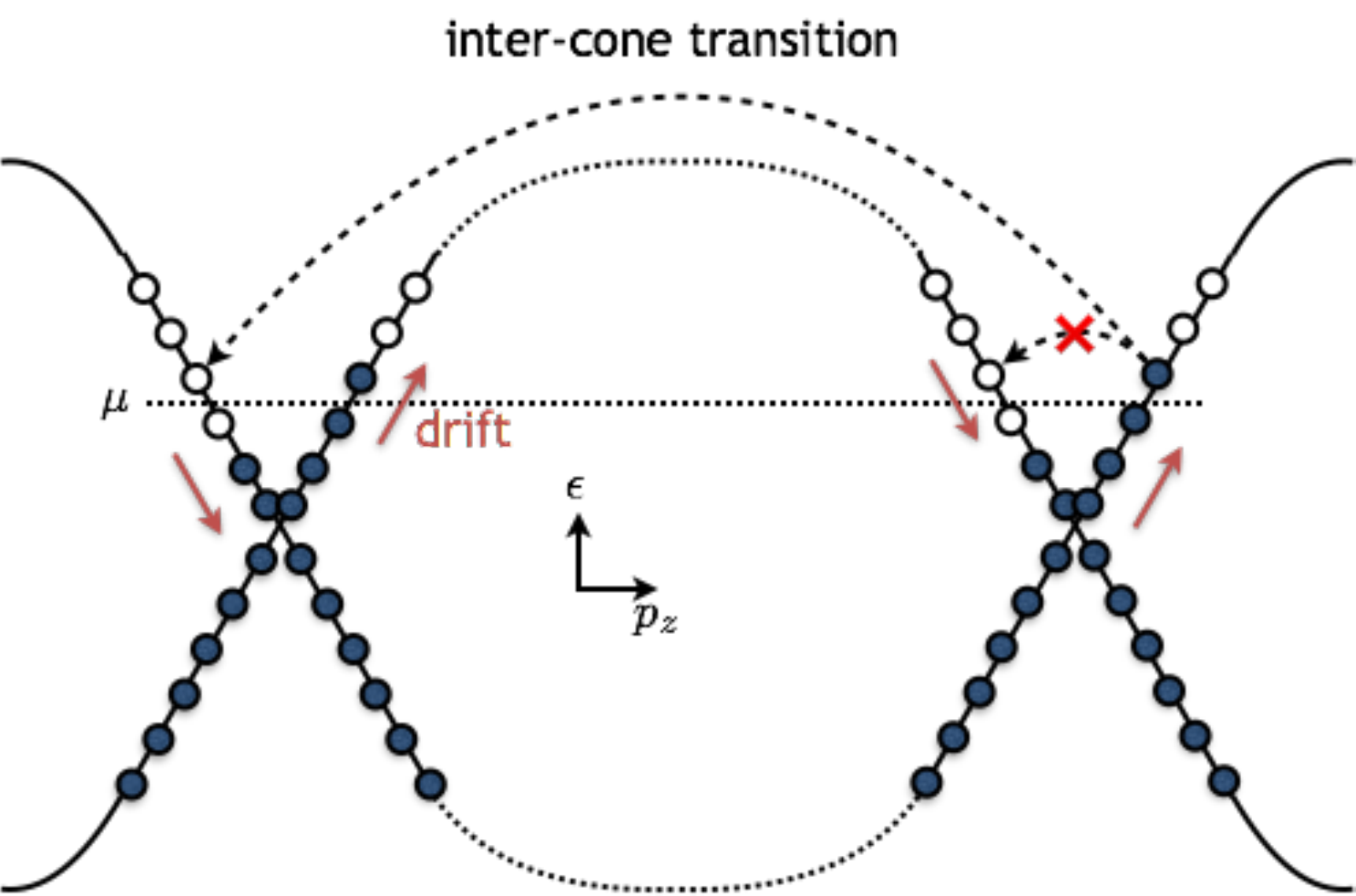}
		\caption{In the inter-cone transition the fermion is scattered into the other cone. Since the inter-cone transition is highly dependent on its modeling, it is difficult to give a universal prediction.}
		\label{inter-cone}
	\end{minipage}
	\hspace{2mm}
	\begin{minipage}{0.5\hsize}
		\centering
		\includegraphics[bb= 0 0 398 355, width=6cm]{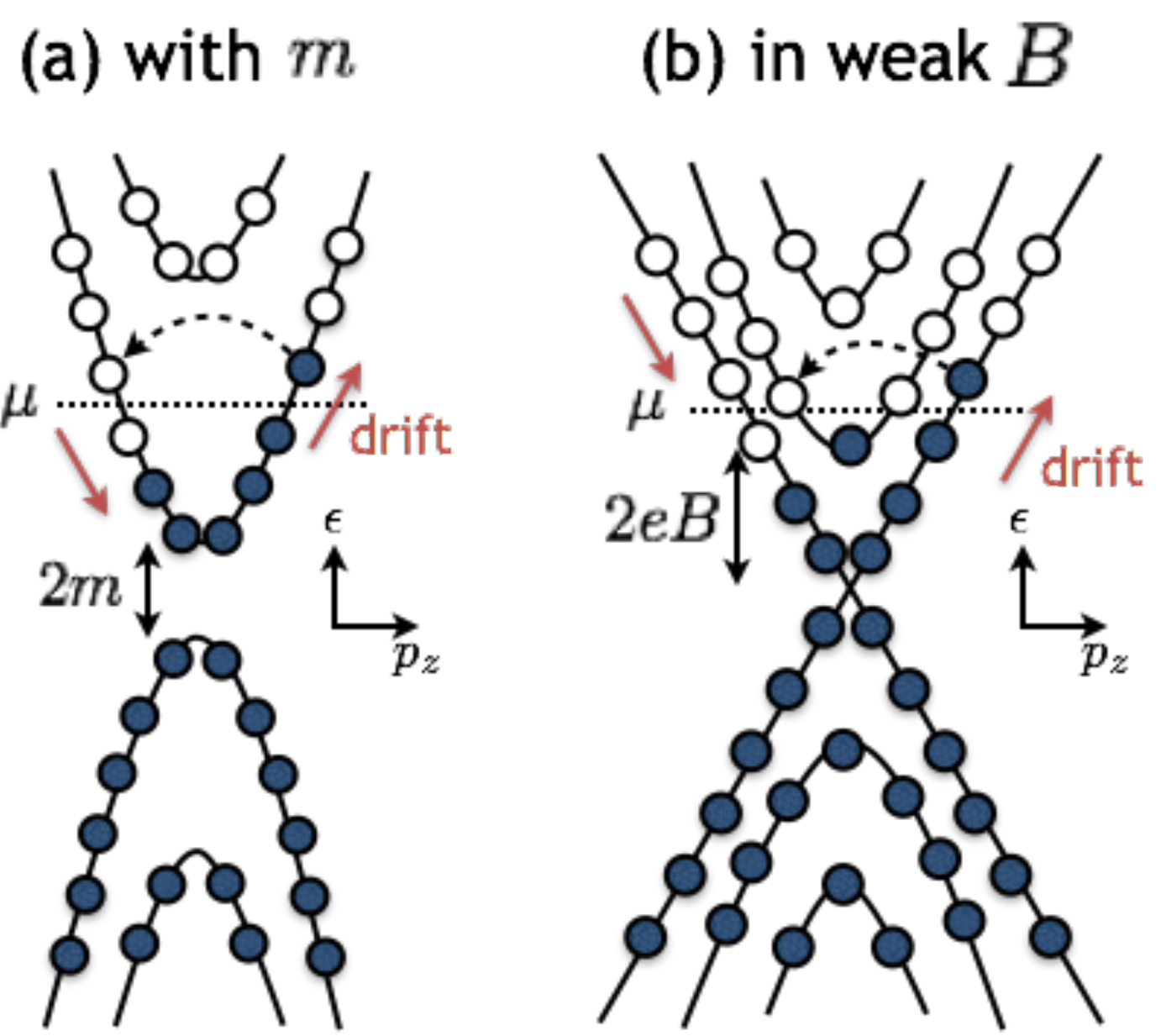}
		\caption{The mechanisms of the intra-cone transition due to (a)the mass and (b)the first excited states in a weak magnetic field region. The effect of the intra-cone transition can be estimated in the low energy effective theory.}
		\label{intra-cone}
	\end{minipage}
\end{figure}

In usual discussions, only the Weyl or Dirac semimetal in the quantum limit are considered. In this case, due to the helicity conservation, only the inter-cone transition take place as shown in Fig.\ref{inter-cone}. In our study, we generalize the situation and consider the electron system with a small mass gap and the magnetic field both in and away from the quantum limit using the low energy effective theory. 

The low energy effective theory description using the relativistic fermion is valid provided that one considers low energy phenomena 
which takes place within the Dirac cone in mind. However,  the transition between two different Dirac cones (inter-cone transition[Fig.\ref{inter-cone}]) cannot be described 
by the low energy theory since the transition amplitude receives non-negligible contributions from the integral over the entire momentum space. 
In the following, we assume that there is always a inter-cone transition which can only be predicted by the full theory, but 
the relaxation time after the onset of the intra-cone transition can be well described by the low energy effective theory. 
Although the situation is different depending on the magnetic field angle, separation of nodes, screening length of the Coulomb impurity and others and should be examined for each setup, we consider the case that the inter-cone transition effect is much smaller than the intra-cone transition[Fig.\ref{intra-cone}] effect. 

We introduce a small mass in order to include a mass gap due to a mechanical strain.
Examples of such a strain-induced mass gap is studied for Bi$_2$Se$_3$ in Ref.\cite{PhysRevB.84.085106}

Consider a probability function $f(n, {\bf p}, t)$ for the electron with the $y$ and $z$ direction momentum ${\bf p}=(p_y, p_z)$ in the $n$-th Landau level. Applying a weak electric field in the same direction as the magnetic field
\begin{eqnarray}
	{\bf E} = ( 0, 0, E), 
\end{eqnarray}
the Boltzmann equation is given by 
\begin{eqnarray}
	\frac{\partial}{\partial t} f(n, {\bf p}, t) - eE\frac{\partial}{\partial p_z} f(n, {\bf p}, t) = \left( \frac{\partial}{\partial t} f(n, {\bf p}, t ) \right)_{\rm coll}, \label{eq:Boltzmann}
\end{eqnarray}
where the second term of the left hand side is the drift term and the right hand side is the collision term. The collision term is defined as
\begin{eqnarray}
	\left( \frac{\partial}{\partial t} f(n, {\bf p}, t ) \right)_{\rm coll}&&= -\sum_{n'} \int _{\rm BZ}\frac{d^2 {\bf p}^\prime}{(2\pi)^2} f(n, {\bf p}, t)W(n, {\bf p}, \rightarrow n^\prime , {\bf p}', )(1- f(n', {\bf p}', t))\nonumber\\
	&&+\sum_{n'} \int _{BZ}\frac{d^2 {\bf p}^\prime}{(2\pi)^2}  f(n', {\bf p}', t)W(n', {\bf p}', \rightarrow n, {\bf p}, )(1 - f(n, {\bf p}, t)) ,
\end{eqnarray}
where $W$ is the transfer probability in unit time. 

Due to the very weak electric field which can be treated as a perturbation, the distribution function is slightly deviated from the equilibrium and can be described as
\begin{eqnarray}
	f(n, {\bf p}, t)=f_0(\epsilon )+\delta f(n, {\bf p}, t),
\end{eqnarray}
where $f_0$ is the probability distribution function in the equilibrium with no electric field and $\delta f$ is the tiny deviation from the equilibrium of $\mathcal{O} (eE)$ for the $n$-th Landau level. $\epsilon $ is the energy of the electron. 

Assuming a small deviation from the equilibrium, the right hand side of the definition of the collision term becomes
\begin{eqnarray}
	\left( \frac{\partial}{\partial t} f(n, {\bf p}, t ) \right)_{\rm coll}= -\sum_{n^\prime } \int _{\rm BZ}\frac{d^2 {\bf p}^\prime}{(2\pi)^2} W(n, {\bf p}\rightarrow n^\prime , {\bf p}')(\delta f(n, {\bf p}, t)- \delta f(n', {\bf p}', t)), 	\label{eq:collision}
\end{eqnarray} 
up to higher order terms in $\delta f$. 

To solve the Boltzmann equation, one often makes the relaxation time approximation, which assume that the probability distribution function exponentially get back into the thermodynamical equilibrium in relaxation time $\tau (n, {\bf p})$ due to the scattering effect: $\delta f\propto e^{-t/\tau }$. Then the collision term can be written as
\begin{align}
	\left( \frac{\partial }{\partial t}f(n, {\bf p}, t)\right) _{\rm coll}=-\frac{\delta f(n, {\bf p}, t)}{\tau (n, {\bf p})}	\label{eq:relaxation}.
\end{align}
Substituting this equation into Eq.(\ref{eq:Boltzmann}), the static solution of the Boltzmann equation is 
\begin{align}
	f(n, {\bf p})=f_0(\epsilon )+eE\tau (n, {\bf p})\frac{\partial f_0(\epsilon )}{\partial p_z}+\mathcal{O}(E^2).	\label{eq:solution_f}
\end{align}
Thus, the deviation from the equilibrium in the lowest order in $eE$ is
\begin{eqnarray}
	\delta f(n, {\bf p})=eE\tau (n, {\bf p})\frac{\partial f_0(\epsilon )}{\partial p_z}=eE\tau (n, {\bf p})\frac{\partial \epsilon (n, {\bf p})}{\partial p_z}f_0'(\epsilon ).	\label{eq:delta_f}
\end{eqnarray}
Substituting Eq.(\ref{eq:relaxation},\ref{eq:delta_f}) into the equation for the definition of the collision term (\ref{eq:collision}) and considering an energy conservation law, the equation to determine the relaxation time $\tau $ is obtained as
\begin{eqnarray}
	\frac{\partial \epsilon(n, {\bf p})}{\partial p_z}&=& \sum_{n'} \int _{\rm BZ}\frac{d^2 {\bf p}^\prime}{(2\pi)^2} W(n, {\bf p}\rightarrow n^\prime , {\bf p}')\nonumber \\
	&&\times \left( \tau(n, {\bf p})  \frac{\partial \epsilon(n, {\bf p})}{\partial p_z} -\tau(n^\prime, {\bf p}') \frac{\partial \epsilon(n', {\bf p}')}{\partial p^\prime_z}  \right) .	\label{eq:tau_eq}
\end{eqnarray}

Only the states around the Dirac points contribute to the low energy physics. Therefore, we can replace the momentum integral in the Brillouin zone:
\begin{eqnarray}
	\int _{\rm BZ}\frac{d^2p}{(2\pi )^2}F({\bf p}),
\end{eqnarray}
with the low energy momentum integral and the sum of cones labeled by $A$
\begin{eqnarray}
	\sum_{A}\int _{\rm low\ energy }\frac{d^2q}{(2\pi )^2}F({\rm p}_A+{\bf q}),
\end{eqnarray}
where $F({\bf p})$ is any function of ${\bf p}$, and ${\bf p}_A$ is the momentum on the Dirac cone labeled by $A$. Then Eq.(\ref{eq:tau_eq}) can be rewritten by 
\begin{eqnarray}
	&&\frac{\partial \epsilon(n, {\bf p}_A+{\bf q})}{\partial q_z}\nonumber \\
	&=& \sum_{n', A'} \int _{\rm low\ energy}\frac{d^2 {\bf q}^\prime}{(2\pi)^2} W(n, {\bf p}_A+{\bf q}\rightarrow n^\prime , {\bf p}_{A'}+{\bf q}')\nonumber \\
	&\times &\left( \tau(n, {\bf p}_A+{\bf q})  \frac{\partial \epsilon(n, {\bf p}_A+{\bf q})}{\partial q_z} -\tau(n^\prime, {\bf p}_{A'}+{\bf q}') \frac{\partial \epsilon(n', {\bf p}_{A'}+{\bf q}')}{\partial q^\prime_z}  \right) .
\end{eqnarray}
In the case that $A=A'$, it denotes the contribution from the intra-cone transition which will be estimated in this paper. In the case that $A\not= A'$, it denotes the contribution from the inter-cone transition which is 
a model- and situation-dependent quantity.

In the following, we consider the intra-cone transition. Then we omit the cone labeling $A$ and the indication `low energy' in the integral. ${\bf p}$ is taken to be the momentum around the Dirac point ${\bf p}_A$. The action of the low energy effective theory for the Dirac fermion is given by 
\begin{eqnarray}
	S=\int d^4x \overline{\psi }(x)\left[ i\Slash{D}-m+\mu \gamma ^0 \right]  \psi (x),
\end{eqnarray}
where $\Slash{D}=\gamma ^\mu D_\mu =\gamma ^\mu (\partial _\mu -ieA_\mu )$. In Appendix \ref{App:Massive}, we derive the wavefunctions under the magnetic field in the Landau level. From the energy eigenvalue (\ref{Landau}), Eq.(\ref{eq:tau_eq}) is simplified as
\begin{eqnarray}
	p_z &=& \sum_{n'} \int \frac{d^2 {\bf p}^\prime}{(2\pi)^2} W(n, {\bf p}\rightarrow n^\prime , {\bf p}')(  \tau(n, {\bf p}) p_z -   \tau(n^\prime, {\bf p}')  p^\prime_z  ), \label{eq:tau_eq_simple}
\end{eqnarray}
where $n$ is a label for the Landau level. In Ref.\cite{PhysRev.104.900}, the relaxation time for a non-relativistic fermion system is calculated in the strong magnetic field region where only the lowest Landau level ($n=0$) contribute.

The electric current density in the $z$ direction is given by
\begin{align}
	J=-e\sum_{n} \int \frac{d^2{\bf p}}{(2\pi )^2}\frac{\partial \epsilon (n, {\bf p})}{\partial p_z}f(n, {\bf p}). 	\label{eq:current}
\end{align}
%
From Eq.(\ref{eq:solution_f}), the expression of the current becomes
\begin{eqnarray}
	J=\sum_{n} \int \frac{d^2p}{(2\pi )^2} (-e)\frac{\partial \epsilon (n, {\bf p})}{\partial p_z}\left( f_0(\epsilon )-eE\tau (n, {\bf p})\frac{\partial f_0(\epsilon )}{\partial p_z}\right) .
\end{eqnarray}
Due to the translation invariance, it turns out that the relaxation time does not depend on the initial momentum $p_y$, and $p_y$ can be regarded as just a label of the degenerate states. Since the probability distribution function is given by the step function:$f_0=\theta (\mu -\epsilon )$ at zero temperature, it becomes
\begin{eqnarray*}
	f(n, {\bf p})&=&\theta (\mu -\epsilon )-eE\tau (n, p_z)\frac{\partial }{\partial p_z}\theta (\mu -\epsilon )\\
	&=&\theta (\mu -\epsilon )+eE\tau (n, p_z)\delta (\mu -\epsilon )\frac{\partial \epsilon }{\partial p_z}.
\end{eqnarray*}
Then the size of the current is expressed as
\begin{eqnarray}
	J
	&=&\frac{-e^3BE}{(2\pi )^2\mu }\sum_{n}\sum_{P_*} |P_*|\tau (n, P_*),	\label{size of current}
\end{eqnarray}
where $P_*$ is defined as the values of $p_z$ which satisfy $\epsilon (p_z)=\mu $.
\section{Relaxation time for massive fermion in strong magnetic field}	\label{sec:Relaxation}
In this section, we give the details of the calculation of the scattering amplitude and the relaxation time for a relativistic fermion with a small mass in the strong magnetic field region. Although there are the inter-cone transition contributions to the relaxation time in the Dirac semimetals, we only consider the scattering within the cone in this paper. 
In the strong magnetic field region only the lowest Landau level $(n, \sigma _3)=(0, +1)$ contribute to the scattering. Since we consider only this state, let us omit the label $(n, \sigma _3)$ in this section. 

We find $P_*$ in Eq.(\ref{size of current}) is
\begin{eqnarray}
	P_1&=&\sqrt{\mu ^2-m^2}, 	\label{P_1}
\end{eqnarray}
or $-P_1$. At zero temperature,  only these two states at the Fermi level contribute to the scattering. Then the size of the current can be written as 
\begin{eqnarray}
	J=\frac{-e^3BE}{(2\pi )^2\mu }P_1(\tau (P_1)+\tau (-P_1))	\label{note-current in strong}.
\end{eqnarray}

Since the probability distributions $f$ and $f_0$ are normalized to unity, their difference satisfies
\begin{eqnarray*}
	\int \frac{d^2p}{(2\pi )^2}\delta f(p_y, p_z)=0, 
\end{eqnarray*}
from which one obtains using Eq.(\ref{eq:delta_f})
\begin{eqnarray}
	\tau (P_1)-\tau (-P_1)=0.	\label{note-tau relation in strong}
\end{eqnarray}
From the Fermi's golden rule, the energy is conserved before and after the transition so that the probability can be 
written as
\begin{eqnarray}
	W(p_y,p_z \rightarrow,p^\prime_y, p^\prime_z) \equiv 2\pi \delta(\epsilon (p_z)-\epsilon(p^\prime_z) )\overline{W}(p_y,p_z \rightarrow, p^\prime_y, p^\prime_z).
\end{eqnarray}
We obtain from Eq.(\ref{eq:tau_eq_simple})
\begin{eqnarray}
	P_1 &=& \int \frac{d^2 {\bf p}^\prime}{(2\pi)^2} 2\pi \delta(\epsilon (P_1)-\epsilon(p^\prime_z) )\overline{W}(p_y,P_1\rightarrow, p^\prime_y, p^\prime_z)(  \tau(P_1) P_1 -   \tau(p^\prime_z)  p^\prime_z  )\nonumber \\
	&=& 2\mu \tau (P_1)\int \frac{dp_y'}{2\pi } \overline{W}(p_y,P_1 \rightarrow p^\prime_y, -P_1). 	\label{P_1 in strong}
\end{eqnarray}

A transition rate per unit time is given by
\begin{eqnarray}
	\overline{W}(p_y,p_z\to p_y',p_z')=\sum_{\bf R}|\langle p_y',p_z'|v({\bf \hat{r}-R})|p_y,p_z\rangle |^2, 	\label{note-probability}
\end{eqnarray}
where ${\bf R}$ stands for a position of the impurity, and ${\bf \hat{r}}$ is a position operator for fermions. The interaction between the fermion and the charged impurity is given by the screened Coulomb potential: 
\begin{eqnarray}
	v({\bf x})=\left( \frac{4\pi e^2}{\kappa }\right) \frac{\exp \left( -|{\bf x}|/r_s\right) }{|{\bf x}|},
\end{eqnarray}
where $r_s$ is the screening length and $\kappa $ is the dielectric constant.

Assuming that the impurities are distributed uniformly with the density $N_I$, we can calculate $\overline{W}$ analytically. We give the expression of 
\begin{eqnarray}
	\int \frac{dp_y'}{2\pi }\overline{W}(p_y, P_1\to p_y', -P_1)
\end{eqnarray}
named $w_{14}$ in Appendix \ref{App:Formulae}. From Eq.(\ref{P_1 in strong}) we find
\begin{eqnarray}
	\tau (P_1)=\frac{\mu P_1}{2w_{14}}.	\label{tau_1}
\end{eqnarray}
This equation and Eq.(\ref{w_14}) yield
\begin{eqnarray}
	\frac{1}{\tau }=\frac{8\pi e^4N_I}{\kappa ^2} \frac{m^2}{\mu \sqrt{\mu ^2-m^2}}\frac{1}{4(\mu ^2-m^2)+1/r_s^2}I\left( \frac{1}{2eB}(4(\mu ^2-m^2)+1/r_s^2)\right) .
\end{eqnarray}
Note that the result by Argyres and Adams for the non-relativistic fermion in the strong magnetic field region\cite{PhysRev.104.900} can be reproduced by artificially taking the non-relativistic limit $m\to \mu $. In the massless limit, since $w_{14}$ goes to zero, the relaxation time goes to infinity, but for the inter-cone transition. It reflects the fact that the helicity does not flip during the impurity scattering in the massless case. Therefore the longitudinal magnetoconductivity is remarkably observed in the massless fermion system such as the Weyl or Dirac semimetals. 

\begin{figure}[tbp]
	\centering
	\includegraphics[bb= 0 0 496 328, width=8cm]{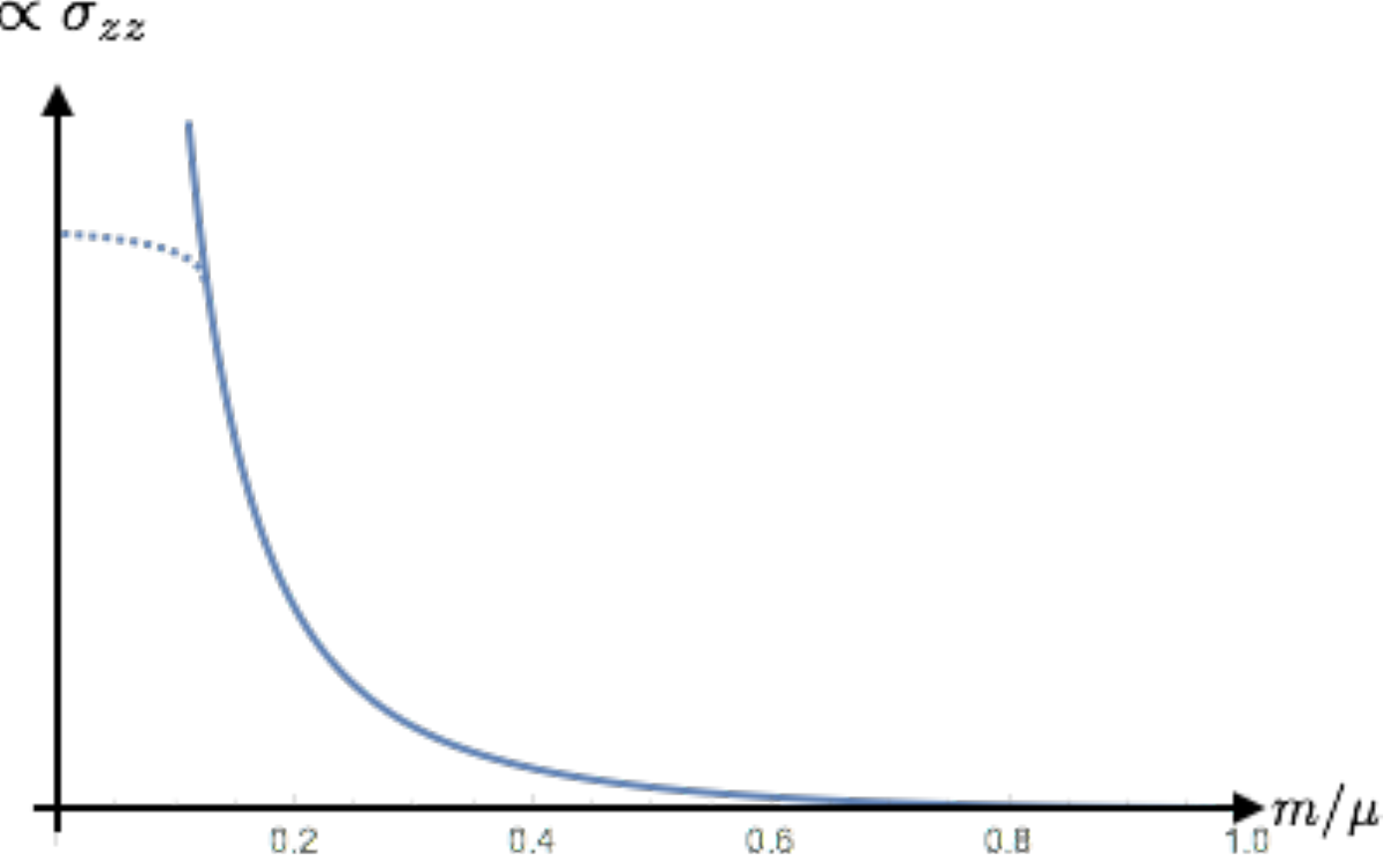}
	\caption{The interpolating function between the relativistic and non-relativistic case for the conductivity: $\sigma _{zz}$ vs. $m/\mu $. The solid line is the estimation without the inter-cone transition. In the massless limit, the conductivity is suppressed by the inter-cone transition (dotted line). }
	\label{Fig:Strong}
\end{figure}

In the strong magnetic field limit $eB\to \infty $, $I\left( \frac{1}{2eB}(4(\mu ^2-m^2)+1/r_s^2)\right) \to 1$, then 
\begin{eqnarray}
	\frac{1}{\tau }\to \frac{8\pi e^4N_I}{\kappa ^2} \frac{m^2}{\mu \sqrt{\mu ^2-m^2}}\frac{1}{4(\mu ^2-m^2)+1/r_s^2}.
\end{eqnarray}
Then we get the interpolating formula between the relativistic and the non-relativistic magnetoconductivities in the strong magnetic field limit as shown in Fig.[\ref{Fig:Strong}]. While $1/\tau $ grows as
\begin{eqnarray}
	\frac{1}{\tau }\propto (\mu ^2-m^2)^{-1/2} \quad (m/\mu \sim 1),
\end{eqnarray}
in the non-relativistic limit, we find 
\begin{eqnarray}
	\frac{1}{\tau }\propto m^2 \quad (m/\mu \sim 0),
\end{eqnarray}
in the relativistic limit. 

Our results show that without the mass gap, there is indeed no intra-cone transition and the scattering is dictated by the inter-cone transition. If the inter-cone transition effect is very small, it gives a large conductivity. Once the mass gap is generated, the intra-cone transition starts to dominate the scattering process,
so that the conductivity reduces for the larger mass gap as $m^{-2}$ . It is interesting to note that for a material like Bi$_2$Se$_3$, there is a mechanism in which the mechanical strain can give a drastic change in the conductivity which may have an interesting technological applications\cite{PhysRevB.84.085106}. 
\section{Magnetoconductivity in weaker magnetic field}	\label{sec:Magnetoconductivity}
Let us consider what happens as we make the magnetic field weaker. As the magnetic field weakens, the energy bands of the excited states lower down. After the first excited states touch the Fermi level, these states open as new channels of the scattering. Similar computations of the transition amplitude between Landau levels are provided in Ref.\cite{1367-2630-18-5-053039,PhysRevB.96.060201}. The calculations in those references are treating the case where the chemical potential is below the threshold so that only off-shell transition appears as an intermediate state in the 2nd order perturbation theory. In our calculation the chemical potential is above the threshold so that the direct transition from the lowest Landau level to the higher Landau level is allowed as the on-shell transition. 

Let us consider the case that only the lowest $(n=0, \sigma _3=+1 )$ and the 1st excited states $(n=0, \sigma _3=-1),\ (n=1, \sigma _3=+1)$ in the Landau levels are below the Fermi energy. Defining $P_1$ and $P_2$ as
\begin{eqnarray}
	P_1&=&\sqrt{\mu ^2-m^2}, 	\label{P_1}\\
	P_2&=&\sqrt{\mu ^2-m^2-2eB}, 	\label{P_2}
\end{eqnarray}
we find that $P_*$ in Eq.(\ref{size of current}) is $\pm P_1$ for the state with $(n=0, \sigma _3=+1)$, and $\pm P_2$ for the states with $(n=1, \sigma _3=+1)$ and $(n=0, \sigma _3=-1)$, respectively. At zero temperature,  only these six set of states (labeled by $I=1, \cdots , 6$) at the Fermi level contributes to the scattering process. We label these states as
\begin{eqnarray*}
	&1=(n=0, \sigma _3=+1, p_z=P_1), 2=(n=1, \sigma _3=+1, p_z=P_2), 3=(n=1, \sigma _3=+1, p_z=-P_2), \nonumber \\
	&4=(n=0, \sigma _3=+1, p_z=-P_1), 5=(n=0, \sigma _3=-1, p_z=P_2), 6=(n=0, \sigma _3=-1, p_z=-P_2).
\end{eqnarray*}
We denote the corresponding relaxation time $\tau _I(I=1, \cdots , 6)$ as:
\begin{eqnarray}
	\tau _1&=&\tau _{0,+}(P_1),\ \tau _2=\tau _{1,+}(P_2),\ \tau _3=\tau _{1,+}(-P_2),\nonumber \\
	\tau _4&=&\tau _{0,+}(-P_1),\ \tau _5=\tau _{0,-}(P_2),\ \tau _6=\tau _{0,-}(-P_2). 
\end{eqnarray}
Then from Eq.(\ref{size of current}), the size of the current can be written as 
\begin{eqnarray}
	J=\frac{-e^3BE}{(2\pi )^2\mu }[P_1(\tau _1+\tau _4)+P_2(\tau _2+\tau _3+\tau _5+\tau _6)]	\label{note-current2}.
\end{eqnarray}

From the Fermi's golden rule, the energy is conserved before and after the transition so that the probability can be 
written as
\begin{eqnarray}
	&&W(n, \sigma _3, p_y,p_z \rightarrow,n^\prime, \sigma_3', p^\prime_y, p^\prime_z) \nonumber \\
	&&\equiv 2\pi \delta(\epsilon (n, \sigma _3, p_z)-\epsilon(n', \sigma _3', p^\prime_z) )\overline{W}(n,\sigma _3, p_y,p_z \rightarrow,n^\prime, \sigma _3', p^\prime_y, p^\prime_z).
\end{eqnarray}
Defining $w_{IJ}$ as 
\begin{eqnarray}
w_{IJ} = \int \frac{dp^\prime_y}{2\pi} \overline{W}(n_I, \sigma _{3I}, p_y, P_I\rightarrow, n_J, \sigma _{3J}, p^\prime_y, P_J) , 
\end{eqnarray}
From a consideration of symmetries, one expects
\begin{eqnarray}
	&w_{IJ}=w_{JI},\quad (I,J=1,\cdots ,6)\nonumber \\
	&w_{12}=w_{43},\ w_{13}= w_{42},\ w_{15}=w_{46},\ w_{16}=w_{54},\ w_{25}=w_{63},\ w_{26}=w_{53}.	\label{note-w_ij relation}
\end{eqnarray}
We compute $w_{IJ}\ (I=1, \cdots , 6)$ using the low energy effective theory in Appendix \ref{App:Formulae}. 

Integrating over $p^\prime_y, p^\prime_z$, we obtain from Eq.(\ref{eq:tau_eq_simple})
\begin{eqnarray}
	P_I=\sum_{J=1}^6w_{IJ}(\tau _IP_I-\tau _JP_J)\frac{\mu }{|P_J|}. 	\label{note-P_i relation}
\end{eqnarray}
Since the probability distributions $f$ and $f_0$ are normalized to unity, their difference satisfies
\begin{eqnarray*}
	\sum_{n, \sigma _3}\int \frac{d^2p}{(2\pi )^2}\delta f(n, \sigma _3, p_y, p_z)=0, 
\end{eqnarray*}
from which one obtains using Eq.(\ref{eq:delta_f})
\begin{eqnarray}
	\tau _1+\tau _2+\tau _5-\tau _3-\tau _4-\tau _6=0.	\label{note-tau relation}
\end{eqnarray}
Combining Eqs.(\ref{note-P_i relation}), the relation of $w_{IJ}$ (\ref{note-w_ij relation}),  and the relation between the relaxation time (\ref{note-tau relation}), we find that the relaxation time on the same bands are equivalent: 
\begin{eqnarray}
	\tau _1=\tau _4,\ \tau _2=\tau _3,\ \tau _5=\tau _6,
\end{eqnarray}
and also obtain simultaneous equations 
\begin{eqnarray}
	1&=&\frac{\mu }{v}\left[ \tau _1\left\{ (w_{12}+w_{13}+w_{15}+w_{16})\frac{1}{P_2}+2w_{14}\frac{1}{P_1}\right\} \right. \nonumber \\
	&&\quad\left. -\tau _2(w_{12}-w_{13})\frac{1}{P_1}-\tau _5(w_{15}-w_{16})\frac{1}{P_1}\right] 	\label{simultaneous1}\\
	1&=&\frac{\mu }{v}\left[ -\tau _1(w_{12}-w_{13})\frac{1}{P_2}+\tau _2\left\{ (w_{12}+w_{13})\frac{1}{P_1}+(2w_{23}+w_{25}+w_{26})\frac{1}{P_2}\right\} \right. \nonumber \\
	&&\quad\left. -\tau _5(w_{25}-w_{26})\frac{1}{P_2}\right] \label{simultaneous2}\\
	1&=&\frac{\mu }{v}\left[ -\tau _1(w_{15}-w_{16})\frac{1}{P_2}-\tau _2(w_{25}-w_{26})\frac{1}{P_2}\right. \nonumber \\
	&&\quad\left. +\tau _5\left\{ (w_{15}+w_{16})\frac{1}{P_1}+(w_{25}+w_{26}+2w_{56})\frac{1}{P_2}\right\} \right] \label{simultaneous3}.
\end{eqnarray}
Solving Eqs.(\ref{simultaneous1}-\ref{simultaneous3}), we can determine the relaxation time in terms of transfer probabilities. The solution of these equations are given in Appendix \ref{App:Solution}. 

\begin{figure}[htbp]
	\centering
	\includegraphics[bb= 0 0 419 272, width=10cm]{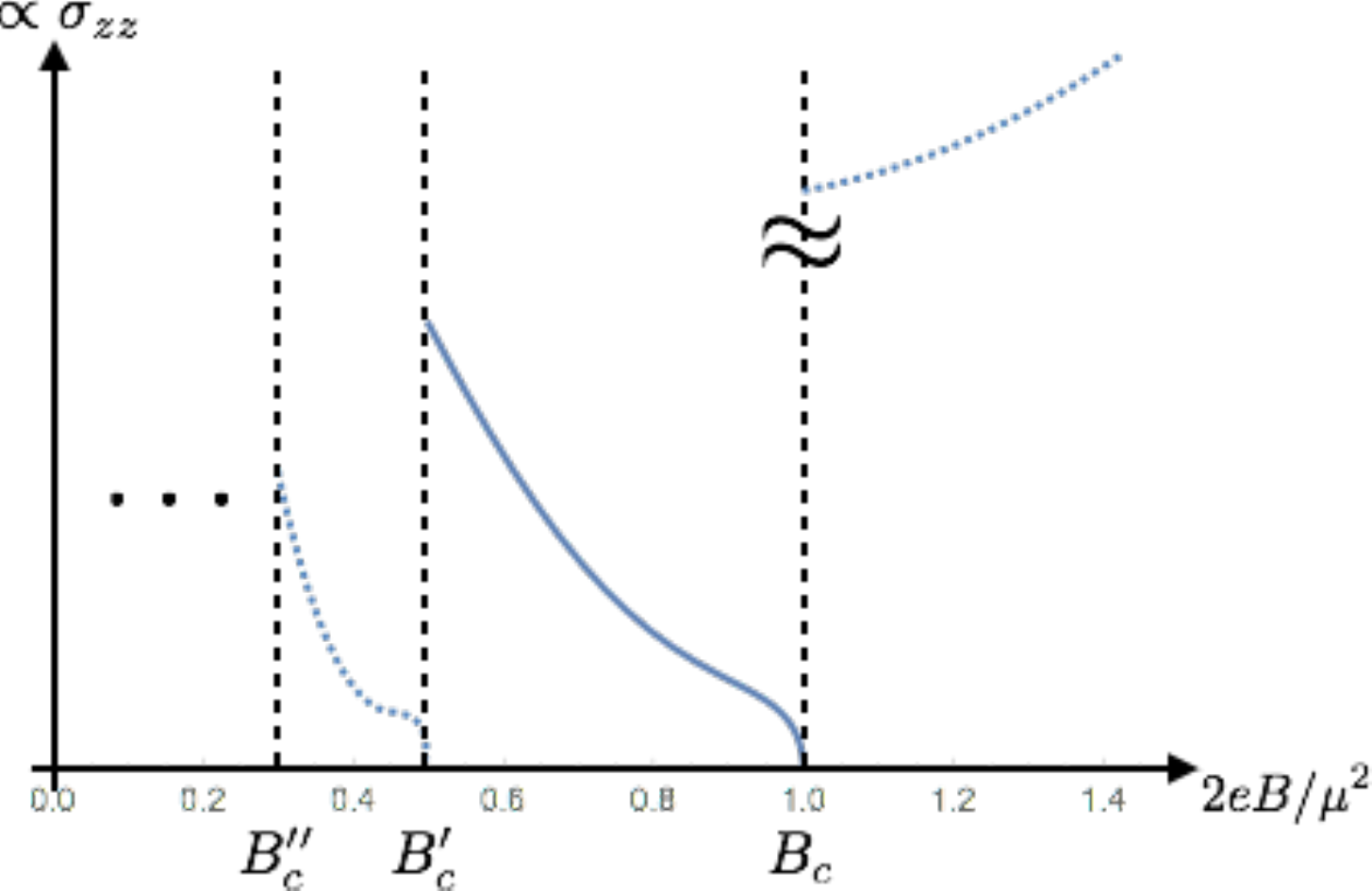}
	\caption{The magnetic field dependence of the conductivity: $\sigma _{zz}$ vs. $2eB/\mu ^2$. The solid line is our result, where our computation is safely applied in the range $B_c'<B<B_c$. When $B>B_c$, since the intra-cone transition is forbidden for a massless fermion, only the inter-cone transition contribute to the suppression of the conductivity. When $B< B_c'$, the second excited states contribute to the intra-cone transition. }
	\label{Fig:Weak}
\end{figure}

Let us make a remark on the gapless case $m=0$.
In Eqs.~(\ref{simultaneous1}), (\ref{simultaneous2}), (\ref{simultaneous3}), the quantity $w_{14}$ vanishes but all other $w_{ij}$'s  do no vanish
and there remains  a nontrivial solution for the relaxation time $\tau_1, \cdots,  \tau_6$. This means that for the magnetic field weaker 
than a critical value $B_{\rm c}=\frac{\mu ^2}{2e}$ where the higher Landau level contributes to the transition, there is a process where the left-handed 
mode is scattered to the right-handed mode in the same light-cone mediated by the higher Landau level states even in the massless limit.  
This effect drastically reduces the conductivity for $B< B_{\rm c}$[Fig.\ref{Fig:Weak}]. For $B<B_{\rm c}'$, where the second excited states contribute to the transition, further reduction of the conductivity is expected. 
\section{Summary and disscussion}	\label{sec:Summary}
We have studied the relaxation time for the relativistic fermions scattered by the Coulomb impurities away from the strong magnetic field limit. We have derived the equations for the relaxation time due to the intra-cone transition, starting from the Boltzmann equation in the relaxation time approximation. Combining with the result of calculation of the transition probability, we have obtained the magnetic field dependence of the longitudinal magnetoconductivity.

In the strong magnetic field region $2eB>\mu ^2-m^2$, we have computed the effect of mass to the intra-cone transition and obtained the interpolating formula between the relativistic and non-relativistic magnetoconductivity. In the massless limit, the relaxation time diverge, because the intra-cone transition does not occur due to the helicity conservation, but for the inter-cone transition. In the non-relativistic limit, our result coincide with the previous result calculated by Argyres and Adams \cite{PhysRev.104.900}. 

In the weak magnetic field region $2eB<\mu ^2-m^2$, the property of the relaxation time drastically changes. The first excited states open as new channels of the scattering. Even in the massless limit, the intra-cone transition through these states, which does not occur in the strong field region, contribute to the finite relaxation times. We have found that as the magnetic field becomes stronger, the conductivity becomes smaller. 

At the border of the strong and weaker magnetic field region where the first excited states just touch the Fermi level, one finds $2eB_{\rm c}=\mu ^2-m^2$. In this case Fig.[\ref{Fig:Weak}] shows that the conductivity goes to zero. This means that at the border of the strong and weak magnetic field region the current goes to zero, while the current becomes very large when the magnetic field exceed that point because only the 
inter-cone transition contribute to the suppression of the conductivity. This phenomenon could be an interesting signal for the longitudinal magnetoconductivity related to the chiral anomaly in the condensed matter system. 

In our study, we considered only the lowest and the first excited bands. When $2eB< \frac{\mu ^2-m^2}{2}$, the second excited states come to contribute to the scattering. So the higher energy band states should be included, when one considers weaker magnetic field case. We considered zero temperature case where scattering by the acoustic phonon can be neglected. However in finite temperature case, we should include the effect of the phonon scattering. Note that the simultaneous equations (\ref{note-P_i relation}) hold when one considers higher excited states or different kinds of scattering sources. Finally, the inter-cone transition and the effect through the surface states \cite{PhysRevB.93.245304} should contribute to the relaxation, which is highly dependent on the materials and its lattice models. The effect of the electron-electron scattering and the longitudinal magnetotransport for Weyl semimetal with multi-monopole has been investigated in Ref.\cite{Boyda:2017dml} and Ref.\cite{PhysRevB.94.195144}, respectively.

\acknowledgments
The work of A. K. was supported in part by the JSPS Research Fellowship for Young Scientists.

\appendix
\section{Massive Dirac fermion in the magnetic field}	\label{App:Massive}
We solve the massive Dirac equation derived from the action
\begin{eqnarray}
	S=\int d^4x \overline{\psi }(x)\left[ i\Slash{D}-m+\mu \gamma ^0 \right]  \psi (x),
\end{eqnarray}
under a magnetic field, where $\Slash{D}=\gamma ^\mu D_\mu =\gamma ^\mu (\partial _\mu -ieA_\mu )$. Here, $\gamma $ matrices are taken to be the Weyl representation: 
\begin{eqnarray}
	\gamma ^\mu =\left( 
	\begin{array}{cc}
		0 & \sigma ^\mu  \\
		\overline{\sigma }^\mu  & 0
	\end{array}
	\right) ,\gamma _5=\left( 
	\begin{array}{cc}
		-1 & 0 \\
		0 & 1
	\end{array}
	\right) .
\end{eqnarray}

We consider the Dirac fermion with the mass $m\not= 0$, which obeys the Dirac equation:
\begin{align}
	\left[ i\Slash{D}-m+\mu \gamma ^0 \right]  \psi (x)=0. 
\end{align}
Since we consider a constant background magnetic field along the $z$ axis, the gauge is taken so that the vector potential corresponding the magnetic field whose magnitude is $B$ is
\begin{eqnarray}
	{\bf A} = ( 0,  Bx, 0 ). 
\end{eqnarray}
In this gauge, the momentum $p_y$ is a good quantum number to label the states in order to distinguish the degenerate states in the $n$-th Landau level. Multiplying $\gamma ^0$ to the Dirac equation from the left, 
\begin{align}
	\left [i\partial _0+i\gamma ^0\gamma ^iD_i-m\gamma ^0\right] \psi (x)=0.	\label{Dirac}
\end{align}
To obtain the wave function $\psi (x)$, first we define the auxiliary function $\Phi $ by
\begin{align}
	\left [i\partial _0+i\gamma ^0\gamma ^iD_i-m\gamma ^0\right] \left[ i\partial _0 -i\gamma ^0\gamma ^iD_i+m\gamma ^0\right] \Phi (x)=0.	\label{Auxiliary}
\end{align}
Then we can get the wave function $\psi (x)$ as
\begin{align}
	\psi (x)=\left[ i\partial _0 -i\gamma ^0\gamma ^iD_i+m\gamma ^0\right] \Phi .
\end{align}
From eq.(\ref{Auxiliary}), the eigenfunction with the energy $\epsilon $ and the momentum in the $y$ and $z$ direction $p_y,\ p_z$ satisfies a harmonic oscillator type equation
\begin{align}
	\left[ -\partial _1^2+(eB)^2\left( x-\frac{p_y}{eB}\right) ^2+p_z^2+m^2-eB\sigma _3\right] \Phi =\epsilon ^2\Phi (x).
\end{align}
Therefore, the energy levels are given by Landau levels
\begin{align}
	\epsilon _{n,\sigma _3}(p_z)=\pm \sqrt{2eB\left( n+\frac{1}{2}\right) +p_z^2+m^2-eB\sigma _3},	\label{Landau}
\end{align}
where each states are degenerated in the $p_y$ space. Note that the states with $(n, \sigma _3=-1)$ and $(n+1, \sigma _3=+1)$ are degenerated except for the lowest Landau level $(n=0, \sigma _3=+1)$. The negative energy states do not contribute to the scattering, because these are always occupied. 

The eigen function $\Phi $ is
\begin{align}
	\Phi _{n,\sigma^3,\gamma _5}(x,p_y,p_z)&=N_{n,\sigma _3, \gamma _5}(p_z)\exp(ip_yy+ip_zz)\sqrt[4]{\frac{eB}{\pi 2^{2n}n!}}\nonumber \\
	&\qquad \times \exp \left[ -\frac{1}{2}eB\left( x-\frac{p_y}{eB}\right) ^2\right] H_n\left( \sqrt{eB}x-\frac{p_y}{\sqrt{eB}}\right) \chi _{\sigma _3, \gamma _5},	\label{eigen function}
\end{align}
where $N_{n,\sigma _3, \gamma _5}(p_z)$ is the normalization constant, $H_n\left( \sqrt{eB}x-\frac{p_y}{\sqrt{eB}}\right) $ is the Hermite polynomials, and $\chi _{\sigma _3, \gamma _5}$ is the eigen spinor of $\sigma _3$, and $\gamma _5$ given by 
\begin{align}
	\chi _{+,L}=\left( 
	\begin{array}{c}
		1\\
		0\\
		0\\
		0
	\end{array}
	\right) ,\chi _{-,L}=\left( 
	\begin{array}{c}
		0\\
		1\\
		0\\
		0
	\end{array}
	\right) ,\chi _{+,R}=\left( 
	\begin{array}{c}
		0\\
		0\\
		1\\
		0
	\end{array}
	\right) ,\chi _{-,R}=\left( 
	\begin{array}{c}
		0\\
		0\\
		0\\
		1
	\end{array}
	\right) .
\end{align}
$\sqrt[4]{\frac{eB}{\pi 2^{2n}n!}}$ is part of the normalization constant, set as 
\begin{align*}
	\int dx\Phi ^\dag \Phi =|N_{n, \sigma _3}(p_z)|^2\chi ^\dag \chi .
\end{align*}
The normalization constant $N_{n,\sigma _3}(p_z)$ is determined by the normalization condition
\begin{align}
	\langle n',\sigma _3';p_y'.p_z'|n,\sigma _3;p_y,p_z\rangle =(2\pi )^2\delta (p_y-p_y')\delta (p_z-p_z')\delta _{n,n'}\delta _{\sigma _3, \sigma _3'}.	\label{note-normalization condition}
\end{align}
\section{The formulae of $w_{IJ}$}	\label{App:Formulae}
\begin{figure}[htbp]
	\centering
	\includegraphics[bb=0 0 403 197, width=8cm]{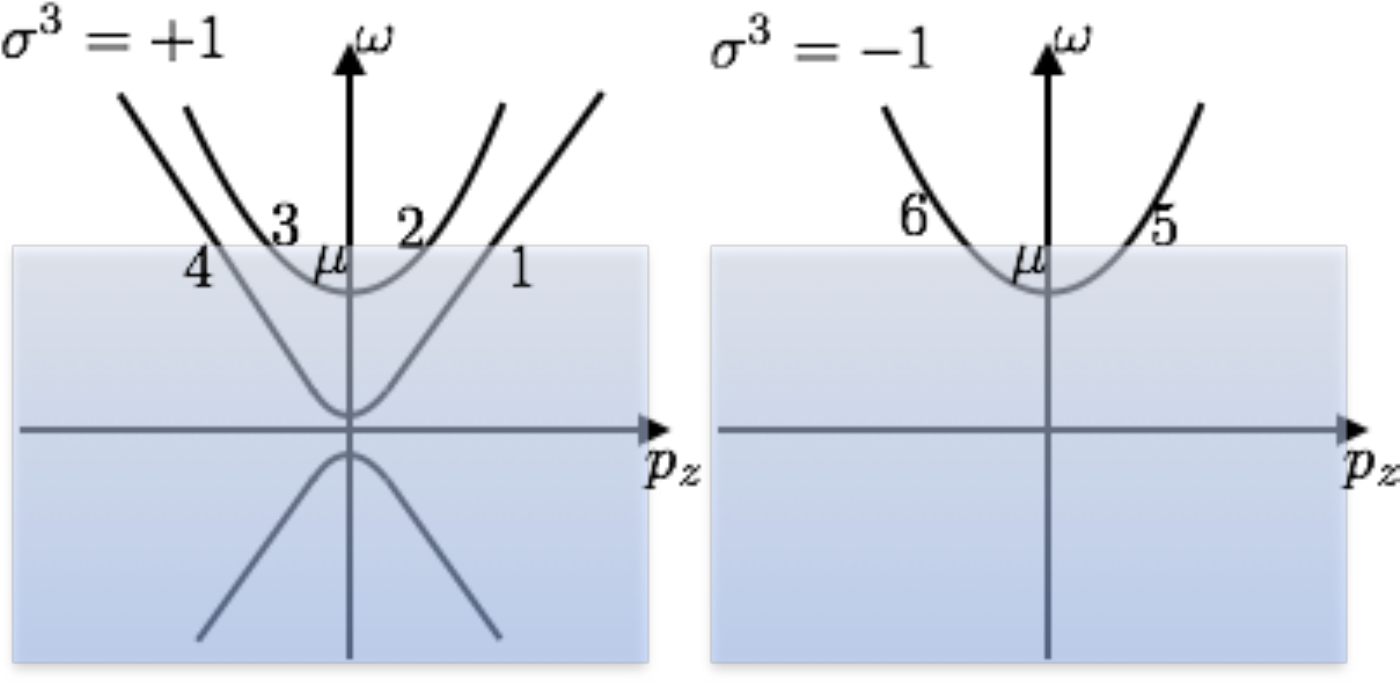}
	\caption{The definition of the labels for the states on the Fermi energy. We keep the same notation even for strong magnetic field case $2eB> \mu ^2-m^2$, in which only the states $1$ and $4$ exist. }
	\label{Fig:IJ definition}
\end{figure}

We give the expression of $w_{IJ}$, which is defined as transition rate per unit time integrated over $p_y'$:
\begin{eqnarray}
	w_{IJ} \equiv \int \frac{dp^\prime_y}{2\pi} \overline{W}(n_I, \sigma _{3I}, p_y, P_I\rightarrow, n_J, \sigma _{3J}, p^\prime_y, P_J) , 
\end{eqnarray}
where $I, J=1,\cdots , 6$ are the labels of the states on the Fermi energy[Fig. \ref{Fig:IJ definition}]. Defining $P_1$ and $P_2$ as
\begin{eqnarray}
	P_1&=&\sqrt{\mu ^2-m^2}, 	\label{P_1}\\
	P_2&=&\sqrt{\mu ^2-m^2-2eB}, 	\label{P_2}
\end{eqnarray}
we find that 
only 
the six set of states (labeled by $I=1, \cdots , 6$) at the Fermi level contributes to the scattering process at zero temperature. We label these states as
\begin{eqnarray*}
	&1=(n=0, \sigma _3=+1, p_z=P_1), 2=(n=1, \sigma _3=+1, p_z=P_2), 3=(n=1, \sigma _3=+1, p_z=-P_2), \nonumber \\
	&4=(n=0, \sigma _3=+1, p_z=-P_1), 5=(n=0, \sigma _3=-1, p_z=P_2), 6=(n=0, \sigma _3=-1, p_z=-P_2).
\end{eqnarray*}
The transition rate per unit time is given by
\begin{eqnarray}
	\overline{W}(p_y,p_z\to p_y',p_z')=\sum_{\bf R}|\langle p_y',p_z'|v({\bf \hat{r}-R})|p_y,p_z\rangle |^2, 	\label{note-probability}
\end{eqnarray}
where ${\bf R}$ stands for a position of the impurity, and ${\bf \hat{r}}$ is a position operator for fermions. The interaction between the fermion and the charged impurity is given by the screened Coulomb potential: 
\begin{eqnarray}
	v({\bf x})=\left( \frac{4\pi e^2}{\kappa }\right) \frac{\exp \left( -|{\bf x}|/r_s\right) }{|{\bf x}|},
\end{eqnarray}
where $r_s$ is the screening length, $\kappa $ is dielectric constant. Similar computations of the transition amplitude between Landau levels are provided in Ref.\cite{1367-2630-18-5-053039,PhysRevB.96.060201}. 

Assuming that the impurities are distributed uniformly with a density $N_I$, we can calculate $w_{IJ}$ analytically. There are nine independent components of $w_{IJ}$: $w_{12}$, $w_{13}$, $w_{14}$, $w_{15}$, $w_{16}$, $w_{23}$, $w_{25}$, $w_{26}$, $w_{56}$. At the Fermi energy $\epsilon (n, \sigma _3, p_z)=\mu $, some straightforward manipulations yield 
\begin{eqnarray}
	w_{12}&=&\left( \frac{4\pi e^2}{\kappa }\right) ^2N_I\frac{\left[ \left( \mu +m\right) ^2+P_1P_2\right] ^2}{4\mu ^2(\mu +m)^2}\nonumber \\
	&&\quad\times \frac{1}{4\pi }\frac{1}{2eB}\frac{1}{\gamma (P_1,P_2)}\left[ 1-(1+\gamma (P_1,P_2))I(\gamma (P_1,P_2))\right]	\label{w_12}, 
\end{eqnarray}
\begin{eqnarray}
	w_{13}&=&\left( \frac{4\pi e^2}{\kappa }\right) ^2N_I\frac{\left[ \left( \mu +m\right) ^2-P_1P_2\right] ^2}{4\mu ^2(\mu +m)^2}\nonumber \\
	&&\quad\times \frac{1}{4\pi }\frac{1}{2eB}\frac{1}{\gamma (P_1,-P_2)}\left[ 1-(1+\gamma (P_1,-P_2))I(\gamma (P_1,-P_2))\right] 	\label{w_13},
\end{eqnarray}
\begin{eqnarray}
	w_{14}=\left( \frac{4\pi e^2}{\kappa }\right) ^2N_I\frac{m^2}{\mu ^2}\frac{1}{4\pi }\frac{1}{2eB\gamma (-P_1,P_1)}I(\gamma (-P_1,P_1))	\label{w_14}, 
\end{eqnarray}
\begin{eqnarray}
	w_{15}&=&\left( \frac{4\pi e^2}{\kappa }\right) ^2N_I\frac{4(2eB)P_2^2}{(4\mu )^2(\mu +m)^2}\nonumber \\
	&&\quad\times \frac{1}{4\pi }\frac{1}{2eB\gamma (P_1,P_2)}\left[ 1-(1+\gamma (P_1,P_2))I(\gamma (P_1,P_2))\right] 	\label{w_15}, 
\end{eqnarray}
\begin{eqnarray}
	w_{16}&=&\left( \frac{4\pi e^2}{\kappa }\right) ^2N_I\frac{4(2eB)P_2^2}{(4\mu )^2(\mu +m)^2}\nonumber \\
	&&\quad\times \frac{1}{4\pi }\frac{1}{2eB\gamma (P_1,-P_2)}\left[ 1-(1+\gamma (P_1,-P_2))I(\gamma (P_1,-P_2))\right] 	\label{w_16},
\end{eqnarray}
\begin{eqnarray}
	w_{23}&=&\left( \frac{4\pi e^2}{\kappa }\right) ^2N_I\frac{4}{(4\mu )^2(\mu +m)^2}\nonumber \\
	&&\quad\times \frac{2eB}{4\pi }\left[ \left( \frac{2m(\mu +m)}{2eB}\right) ^2\frac{-2-\gamma (P_2,-P_2)+(3+4\gamma (P_2,-P_2)+\gamma ^2(P_2,-P_2))I(\gamma (P_2,-P_2))}{\gamma (P_2,-P_2)}\right. \nonumber \\
	&&\quad+\left( \frac{2m(\mu +m)}{2eB}\right) \frac{-6-2\gamma (P_2,-P_2)+(10+10\gamma (P_2,-P_2)+2\gamma ^2(P_2,-P_2))I(\gamma (P_2,-P_2))}{\gamma (P_2,-P_2)}\nonumber \\
	&&\quad\left. +\frac{-4-\gamma (P_2,-P_2)+(8+6\gamma (P_2,-P_2)+\gamma ^2(P_2,-P_2))I(\gamma (P_2,-P_2))}{\gamma (P_2,-P_2)} \right] 	\label{w_23},
\end{eqnarray}
\begin{eqnarray}
	w_{25}=\left( \frac{4\pi e^2}{\kappa }\right) ^2N_I\frac{8eBP_2^2}{(4\mu )^2(\mu +m)^2}\frac{-1+(2+\gamma (P_2,P_2))I(\gamma (P_2, P_2))}{4\pi (2eB)}	\label{w_25},
\end{eqnarray}
\begin{eqnarray}
	w_{26}=\left( \frac{4\pi e^2}{\kappa }\right) ^2N_I\frac{4P_2^2}{(4\mu )^2(\mu +m)^2}\frac{(\gamma ^2(P_2, -P_2)+4\gamma (P_2, -P_2)+6)I(\gamma (P_2, -P_2))-\gamma (P_2, -P_2)-2}{4\pi \gamma (P_2, -P_2)},\nonumber \\
	\label{w_26}
\end{eqnarray}
\begin{eqnarray}
	w_{56}&=&\left( \frac{4\pi e^2}{\kappa }\right) ^2N_I\frac{4}{(4\mu )^2(\mu +m)^2}\frac{1}{4\pi (2eB)\gamma (P_2, -P_2)}\left[ \left\{ (2eB)^2\gamma ^2(P_2, -P_2)\right. \right. \nonumber \\
	&+& \left. 4eB(6eB+2m(\mu +m))\gamma (P_2, -P_2)+4\left\{ m(\mu +m)+2eB\right\} \left\{ m(\mu +m)+4eB\right\} \right\} I(\gamma (P_2, -P_2))\nonumber \\
	&-&\left. 2eB\left\{ 2eB\gamma (P_2, -P_2)+4\left( m(\mu +m)+2eB\right) \right\} \right] ,\label{w_56}
\end{eqnarray}
where 
\begin{eqnarray}
	\gamma (p_z, p_z')\equiv \frac{(p_z'-p_z)^2+1/r_s^2}{2eB}, 
\end{eqnarray}
and 
\begin{eqnarray}
	I(\gamma )\equiv \int_0^\infty dx\frac{x }{x+\gamma }e^{-x}=1+\gamma e^\gamma {\rm Ei}(-\gamma ) ,
\end{eqnarray}
${\rm Ei}$ being the exponential integral. 
\section{Solution of the equations for relaxation times}	\label{App:Solution}
Simultaneous equations for the relaxation times (\ref{simultaneous1}-\ref{simultaneous3}) can be written as
\begin{eqnarray}
	\left( 
	\begin{array}{c}
		1\\
		1\\
		1
	\end{array}
	\right) =\frac{\mu }{v}\left( 
	\begin{array}{ccc}
		M_{11} & M_{12} &M_{13} \\
		M_{21} & M_{22}&M_{23}\\
		M_{31} & M_{32}&M_{33}
	\end{array}
	\right) \left( 
	\begin{array}{c}
		\tau _1\\
		\tau _2\\
		\tau _5
	\end{array}
	\right) ,
\end{eqnarray}
where 
\begin{eqnarray}
	M_{11}&=&(w_{12}+w_{13}+w_{15}+w_{16})\frac{1}{P_2}+2w_{14}\frac{1}{P_1}, \\
	M_{12}&=&-(w_{12}-w_{13})\frac{1}{P_1}, \\
	M_{13}&=&-(w_{15}-w_{16})\frac{1}{P_1},\\
	M_{21}&=&-(w_{12}-w_{13})\frac{1}{P_2},\\
	M_{22}&=&(w_{12}+w_{13})\frac{1}{P_1}+(2w_{23}+w_{25}+w_{26})\frac{1}{P_2},\\
	M_{23}&=&-(w_{25}-w_{26})\frac{1}{P_2},\\
	M_{31}&=&-(w_{15}-w_{16})\frac{1}{P_2},\\
	M_{32}&=&-(w_{25}-w_{26})\frac{1}{P_2}, \\
	M_{33}&=&(w_{15}+w_{16})\frac{1}{P_1}+(w_{25}+w_{26}+2w_{56})\frac{1}{P_2}.
\end{eqnarray}
Deriving the inverse matrix of $M$, we obtain the relaxation times 
\begin{eqnarray}
	\tau _1&=&\frac{v}{\mu }\frac{M_{13}M_{22}-M_{12}M_{23}-M_{13}M_{32}+M_{23}M_{32}+M_{12}M_{33}-M_{22}M_{33}}{M_{13}M_{22}M_{31}-M_{12}M_{23}M_{31}-M_{13}M_{21}M_{32}+M_{11}M_{23}M_{32}+M_{12}M_{21}M_{33}-M_{11}M_{22}M_{33}}\nonumber \\
	\\
	\tau _2&=&\frac{v}{\mu }\frac{-M_{13}M_{21}+M_{11}M_{23}+M_{13}M_{31}-M_{23}M_{31}-M_{11}M_{33}+M_{21}M_{33}}{M_{13}M_{22}M_{31}-M_{12}M_{23}M_{31}-M_{13}M_{21}M_{32}+M_{11}M_{23}M_{32}+M_{12}M_{21}M_{33}-M_{11}M_{22}M_{33}}\nonumber \\
	\\
	\tau _5&=&\frac{v}{\mu }\frac{M_{12}M_{21}-M_{11}M_{22}-M_{12}M_{31}+M_{22}M_{31}+M_{11}M_{32}-M_{21}M_{32}}{M_{13}M_{22}M_{31}-M_{12}M_{23}M_{31}-M_{13}M_{21}M_{32}+M_{11}M_{23}M_{32}+M_{12}M_{21}M_{33}-M_{11}M_{22}M_{33}}. \nonumber \\
\end{eqnarray}









\bibliographystyle{JHEP}	
\providecommand{\href}[2]{#2}\begingroup\raggedright\endgroup

\end{document}